\documentclass{osa-article}
\journal{oe}

\begin{document}

\title{Chiral photon blockade in the spinning Kerr resonator}

\author{Yunlan Zuo,\authormark{1,5}, Ya-Feng Jiao,\authormark{2,3,5}, Xun-Wei Xu,\authormark{1}, Adam Miranowicz,\authormark{4}, Le-Man Kuang,\authormark{1,2,*} and Hui Jing\authormark{1,2,$\dagger$}}

\address{\authormark{1}Key Laboratory of Low-Dimensional Quantum Structures and Quantum Control of Ministry of Education, Department of Physics and Synergetic Innovation Center for Quantum Effects and Applications, Hunan Normal University, Changsha 410081, China}
\address{\authormark{2}Academy for Quantum Science and Technology, Zhengzhou University of Light Industry, Zhengzhou 450002, China}
\address{\authormark{3}School of Electronics and Information, Zhengzhou University of Light Industry, Zhengzhou 450001, China}
\address{\authormark{4}Institute of Spintronics and Quantum Information, Faculty of Physics, Adam Mickiewicz University, 61-614 Pozna$\acute{\textsl{n}}$, Poland}
\address{\authormark{5}Co-first authors with equal contribution.}

\email{\authormark{*}lmkuang@hunnu.edu.cn}
\email{\authormark{$\dagger$}jinghui73@foxmail.com} 
\begin{abstract}
We propose how to achieve chiral photon blockade by spinning a nonlinear optical resonator. We show that by driving such a device at a fixed direction, completely different quantum effects can emerge for the counter-propagating optical modes, due to the spinning-induced breaking of time-reversal symmetry, which otherwise is unattainable for the same device in the static regime. Also, we find that in comparison with the static case, robust non-classical correlations against random backscattering losses can be achieved for such a quantum chiral system. Our work, extending previous works on the spontaneous breaking of optical chiral symmetry from the classical to purely quantum regimes, can stimulate more efforts towards making and utilizing various chiral quantum effects, including applications for chiral quantum networks or noise-tolerant quantum sensors.
\end{abstract}

\section{Introduction}
Chirality, a mirror symmetry widely existing in nature, plays an essential role in modern science and technology~\cite{Lodahl2017Chiral}. In particular, optical chirality can serve as a unique degree of freedom to engineer light-matter interactions, allowing for applications in e.g., unidirectional-microlasers~\cite{Peng2016Chiral,Song2010Directional,Wong2016Lasing}, exceptional single-photon emissions~\cite{Huang2022Exceptional,Tang2019On,Dong2021All,Bo2023Spinning,Lu2023Chiral}, and enhanced optical gyroscopes~\cite{Ge2015Rotation,Kim2015Non} or nanoparticle sensors~\cite{Wiersig2014Enhancing,Wiersig2016Sensors}. In a recent experiment, the spontaneous breaking of chiral symmetry in the classical domain, characterized by a transition from symmetric bidirectional optical transmissions to a one-way optical flow, was demonstrated by driving a nonlinear resonator beyond a critical point of a pump power~\cite{Experimental2017Cao}. Besides, the chiral absorption via backscattering can be achieved experimentally by adjusting the phase difference of two counterpropagating driving fields in a single resonator~\cite{Ren2023Backscattering}. Other ways to achieve classical optical chiral effects include e.g., inducing non-Hermitian phase transitions~\cite{Peng2016Chiral} or exploiting optical spin-orbit coupling~\cite{Wang2009Observation,Petersen2014Chiral,Bliokh2008Geometrodynamics,Bliokh2015Spin}. However, the possible emergence of highly asymmetric quantum optical correlations in a single nonlinear resonator, as far as we know, has not been explored till now, hindering its potential applications in chiral quantum engineering.

Here we propose how to achieve chiral photon blockade by spinning a nonlinear optical resonator. We note that in a very recent experiment, nonreciprocal propagation of light with $99.6\%$ isolation was realized by spinning a purely optical resonator~\cite{Maayani2018Flying}. The merits of such a spinning device include: continuous tunability, the absence of a power threshold, and easy extension to the quantum domain. Based on spinning systems, nonreciprocal quantum effects i.e., distinct quantum correlations exhibited for output lights when driving the same system from opposite directions, were predicted and soon experimentally confirmed with different systems~\cite{huang2018Nonreciprocal,Jiao2020Nonreciprocal,Li2019Nonreciprocal,Non2023Yang,Nonreciprocity2022Graf,Yuan2023Optical}. Similar spinning techniques were also used to achieve analog thermal materials~\cite{Tunable2020Xu}, quantum droplets~\cite{Rotating2021Dong}, and adaptive thermal cloaking~\cite{Inverse2021Zhu}. However, these works mainly focused on nonreciprocal effects featuring different directions of incident light on the same device, without exploring the exotic possibility to achieve quantum chiral effects for a fixed input light.

\begin{figure*}[t]
    \centering
    \includegraphics[width=1.0\columnwidth]{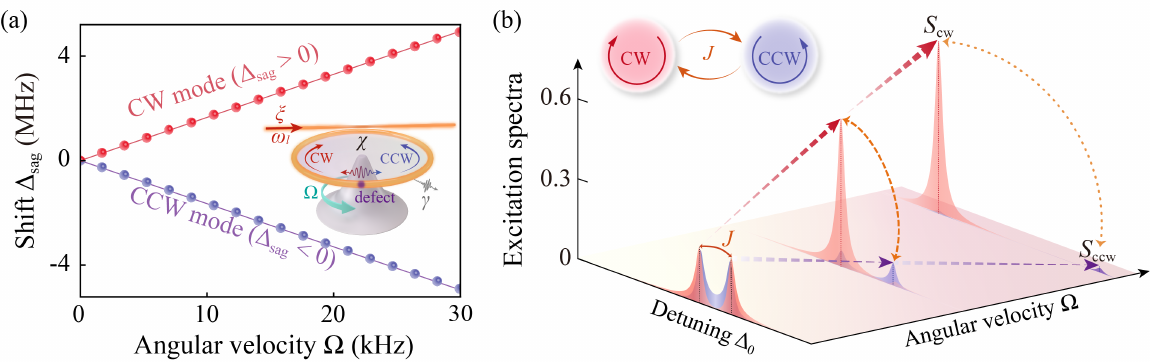}
    \caption{{Spinning-induced optical chirality in a resonator with backscattering.} {(a)} A Kerr-type nonlinearity optical resonator with backscattering spinning along the CCW direction at an angular velocity $\Omega$. Moreover, imperfections of devices, such as surface roughness or material defects, can cause optical backscattering, as effectively described by the mode-coupling strength $J$ between the CW and CCW modes. The Sagnac-Fizeau shift $\Delta_{\mathrm{sag}}$ versus $\Omega$. By fixing the rotation of the resonator, $\Delta_{\mathrm{sag}}>0$ ($\Delta_{\mathrm{sag}}<0$) corresponds to the Sagnac-Fizeau shift of the CW (CCW) mode. {(b)} The cavity excitation spectra $S_{j=\mathrm{cw,ccw}}$ in the breaking of optical chiral symmetry can be observed by tuning various values of $\Omega$. The parameters are given in the second section of the main text. Specifically, in our simulations, we have chosen the values $J/\gamma=2$, $\chi/\gamma=9.5$, $\xi/\gamma=0.25$.}
\label{fig:FP1}
\end{figure*}

Specifically, we consider a spinning Kerr resonator and reveal the possibility of achieving chiral photon blockade. We find that by increasing the angular velocity, optical chirality can emerge not only for classical optical mean numbers of the opposite propagating modes, but also for their quantum correlations. As a result, highly asymmetric photon blockade, a purely quantum effect~\cite{imamoglu1997Strongly,birnbaum2005Photon,rabl2011Photon,lang2011Observation,liao2013Photon,miranowicz2013Twophoton,Kowalewska2019Two}, can emerge only for one mode but not for the other one. Also, we find that robustness against random losses and even a coherent switch of photon blockade can be achieved for such a quantum chiral system, by tuning its angular velocity. Our work indicates that experimentally accessible chiral devices cannot only be used to control classical transmission of light, but also serve as a powerful tool to achieve and manipulate quantum chiral effects, with applications in chiral quantum networks~\cite{Pichler2015Quantum,Mahmoodian2016Quantum,Ramos2016Non,Mok2020Microresonators} and backscattering-immune multi-photon blockade~\cite{hamsen2017TwoPhoton,Chakram2022Multimode,Chakram2021Seamless}.

\section{Physical system}

In recent experiments, spinning devices have been used to
realize directional heat flow~\cite{Tunable2020Xu,Li2019Anti},  resonator gyroscope~\cite{Mao2022Experimental}, and
sound isolators~\cite{Fleury2014Sound,Ding2019Experimental}.  Particularly, an optical diode with $99.6\%$ isolation was demonstrated by spinning an optical resonator~\cite{Maayani2018Flying}, without relying on any magnetic materials or complex structures. Spinning systems have been used to predict nonreciprocal quantum correlation effects~\cite{huang2018Nonreciprocal,Jiao2020Nonreciprocal}, which were then demonstrated with a solid-state device~\cite{Nonreciprocity2022Graf} and a cavity QED system~\cite{Non2023Yang} in very recent experiments. However, the novel possibility of achieving quantum chiral effects for a fixed input light by utilizing such a spinning scheme, as far as we know, has not been explored.

As shown in Fig.~\ref{fig:FP1}(a), we consider a spinning whispering-gallery-mode (WGM) resonator with a Kerr-type optical nonlinearity. The WGM can resonator support two counterpropagating optical modes, i.e., the clockwise (CW) and counterclockwise (CCW) modes, which are basically degenerate. However, any non-ideality of the WGM resonator, such as surface roughness or material inhomogeneity, can couple the CW and CCW modes with strength $J$ and lift their degeneracy due to the optical backscattering effect, which is unavoidable due to manufacturing limitations~\cite{Mohageg2007Coherent,Kippenberg2002Modal,Gorodetsky2000Rayleigh,Weiss1995Splitting}. For a low-quality (Q) WGM resonator, optical backscattering usually leads to a broadened linewidth of the spectral response. On the other hand, for a high-Q WGM resonator, there is a visible frequency splitting introduced to the transmission spectra~\cite{Kim2019Dynamic}. When the WGM resonator of a radius $R$ rotates at an angular velocity $\Omega$, the light circulating in this resonator experiences a Sagnac-Fizeau shift, i.e., $\omega_{c}\rightarrow\omega_{c}+\Delta_{\mathrm{sag}}$, with~\cite{Malykin2000The}\label{eq:sag}
\begin{equation}
\Delta_{\mathrm{sag}}=\pm\frac{n_{1}R\Omega\omega_{c}}{c}\left(1-\frac{1}{n_{1}^{2}}-\frac{\lambda}{n_{1}}\frac{dn_{1}}{d\lambda}\right), \end{equation}
where $\omega_{c}$ is the optical resonance frequency for the static case, $c$ $\left(\lambda\right)$ is the speed (wavelength) of light in vacuum, and $n_{1}$ is the linear refractive index of the material. The dispersion term $dn_{1}/d\lambda$, characterizing the relativistic origin of the Sagnac effect, is relatively small in typical materials ($\sim 1\%$)~\cite{Maayani2018Flying,Malykin2000The}. By fixing the CCW rotation of the resonator, hence $\Delta_{\mathrm{sag}}>0$ ($\Delta_{\mathrm{sag}}<0$) corresponds to a Sagnac-Fizeau shift of the CW (CCW) mode, i.e., $\omega_{\mathrm{cw,ccw}}=\omega_{c}\pm\left|\Delta_{\mathrm{sag}}\right|$, as shown in Fig.~\ref{fig:FP1}(a). In the rotating frame at the pump frequency $\omega_{l}$, the Hamiltonian of the spinning system, with a driving field input from the CW direction, read $\left(\hbar=1\right)$~\cite{huang2018Nonreciprocal,Jiao2020Nonreciprocal,Jing2018Nanoparticle}
\begin{align}\label{eq:H}
\hat{H}=& \left(\Delta_{0}+|\Delta_{\mathrm{sag}}|\right)\hat{a}_{\mathrm{cw}}^{\dagger}\hat{a}_{\mathrm{cw}}+\left(\Delta_{0}-|\Delta_{\mathrm{sag}}|\right)\hat{a}_{\mathrm{ccw}}^{\dagger}\hat{a}_{\mathrm{ccw}}\nonumber \\
&+J\left(\hat{a}_{\mathrm{cw}}^{\dagger}\hat{a}_{\mathrm{ccw}}+\hat{a}_{\mathrm{ccw}}^{\dagger}\hat{a}_{\mathrm{cw}}\right)+\chi\left(\hat{a}_{\mathrm{cw}}^{\dagger2}\hat{a}_{\mathrm{cw}}^{2}+\hat{a}_{\mathrm{ccw}}^{\dagger2}\hat{a}_{\mathrm{ccw}}^{2}\right)\nonumber \\
&+2\chi\hat{a}_{\mathrm{cw}}^{\dagger}\hat{a}_{\mathrm{cw}}\hat{a}_{\mathrm{ccw}}^{\dagger}\hat{a}_{\mathrm{ccw}}+\xi\left(\hat{a}_{\mathrm{cw}}^{\dagger}+\hat{a}_{\mathrm{cw}}\right),
\end{align}
where $\Delta_{0}=\omega_{c}-\omega_{l}$ is the optical detuning between the cavity field and the driving field, $a_{\mathrm{cw}}$ ($a_{\mathrm{cw}}^{\dagger}$) and $a_{\mathrm{ccw}}$ ($a_{\mathrm{ccw}}^{\dagger}$ ) are the annihilation (creation) operators of the CW and CCW modes, respectively. The Kerr parameter is $\chi=\hbar\omega_{c}^{2}cn_{2}/n_{1}^{2}V_{\mathrm{eff}}$, where $n_{2}$ is the nonlinear refractive index of material, $V_{\mathrm{eff}}$ is the effective mode volume of the resonator. The Hamiltonian in Eq.~(\ref{eq:H}) contains both self-Kerr and cross-Kerr interaction terms. $\xi=\sqrt{\gamma P_{\mathrm{in}}/\hbar\omega_{l}}$ is the driving amplitude with a cavity loss rate $\gamma$ and a driving power $P_{\mathrm{in}}$.

First, we analyze the cavity excitation spectra,
\begin{equation}
S_{j}\left(\Delta_{0}\right)=\frac{N_{j}}{n_{0}}=\frac{\langle \hat{a}_{j}^{\dagger}\hat{a}_{j}\rangle}{n_{0}}, \quad \left(j=\mathrm{cw,ccw}\right),
\end{equation}
where $n_{0}=4\,\xi^{2}/\gamma^{2}$ is the normalization factor. The cavity excitation spectra $S_{j}\left(\Delta_{0}\right)$ are the rescaled intracavity photon numbers $N_{j}=\langle \hat{a}_{j}^{\dagger}\hat{a}_{j}\rangle$, which can be obtained by solving the Lindblad master equation~\cite{Johansson2012QuTiP,Johansson2013QuTiP}
\begin{equation}\label{eq:MQ}
\dot{\hat{\rho}}=-i\left[\hat{H},\hat{\rho}\right]+\underset{j}{\sum}\frac{\gamma}{2}\left(2\hat{a}_{j}\hat{\rho}\hat{a}_{j}^{\dagger}-\hat{a}_{j}\hat{a}_{j}^{\dagger}\hat{\rho}-\hat{\rho}\hat{a}_{j}\hat{a}_{j}^{\dagger}\right),
\end{equation}
where $\hat{\rho}$ is the reduced density matrix of the system,  $\rho_{\mathrm{ss}}$ is steady-state solutions of the master
equation. In our calculations, the experimentally accessible parameters are given by~\cite{schuster2008nonlinear,vahala2003optical,spillane2005ultrahigh,Pavlov2017Soliton,huet2016millisecond,zielinska2017self}: $\lambda=1550\:\mathrm{nm}$, $Q=5\times10^{9}$, $V_{\mathrm{eff}}=310\:\mu\mathrm{m}^{3}$, $n_{1}=1.4$, $n_{2}=3\times10^{-14}\:\mathrm{m}^{2}/\mathrm{W}$, $P_{\mathrm{in}}=2\:\mathrm{fW}$, $R=30\:\mu\mathrm{m}$. Notably, $V_{\mathrm{eff}}$ is typically $10^{2}-10^{4}\:\mu\mathrm{m}^{3}$~\cite{vahala2003optical,spillane2005ultrahigh}, and $Q$ is typically $10^{9}-10^{12}$~\cite{Pavlov2017Soliton,huet2016millisecond} for the WGM resonator. Also, we have used the dimensionless parameters $\chi/\gamma=9.5$ and $\xi/\gamma=0.25$. In addition, we have set $J/\gamma=2$, which is well within the experimental feasibility. In fact, a recent experiment~\cite{Zhu2010On} shows that such backscattering-induced mode coupling strength can reach up to about $31.5\:\mathrm{MHz}$, corresponding to $J/\gamma\sim15$.

For the static case ($\Omega=0$), we can observe a symmetry split
resonance in the excitation spectrum [i.e., $\mathrm{max}(S_{\mathrm{cw}})=\mathrm{max}(S_{\mathrm{ccw}})=0.21$], where the field intensity distributions of the CW and CCW modes are in balance~[Fig.~\ref{fig:FP1}(b)]. In contrast, for the spinning case ($\Omega\neq0$), such field intensity distributions of $S_{\mathrm{cw}}$ and $S_{\mathrm{ccw}}$ are no longer equal, indicating that the split resonance becomes asymmetric and the chiral symmetry is broken. By further increasing $\Omega$, it is seen that the excitation spectra of the CW mode $S_{\mathrm{cw}}$ is characterized by a pronounced resonance peak [$\mathrm{max}(S_{\mathrm{cw}})=0.65$], whereas such peak almost vanishes for the CCW mode [$\mathrm{max}(S_{\mathrm{ccw}})=0.01$]. Essentially, with the increase of the angular velocity, the frequency difference between the CW and CCW modes is simultaneously amplified, thus diminishing the coupling between these modes. As a result, the photons that scatter from the CW mode to the CCW mode induced by the mode coupling tend to be suppressed when the WGM resonator spins faster, which leads to the asymmetric split resonance. This asymmetric internal field distribution is the defining hallmark of the chiral modes~\cite{Peng2016Chiral,Experimental2017Cao}. Similarly, when fixing the CCW rotation of the resonator and applying the drive field input from the CCW direction, the chiral-symmetry breaking can still be observed in the cavity excitation spectra.

\section{Photon blockade effect}

Now, we further extend our research to a quantum regime, i.e., exploring the influence of such chirality on the photon blockade (PB) effect. The quantum feature of the light can be characterized by the $\mu$th-order correlation function $g^{(\mu)}\left(0\right)=\left\langle\hat{a}^{\dagger\mu}\hat{a}^{\mu}\right\rangle/\left\langle\hat{a}^{\dagger}\hat{a}\right\rangle^{\mu}$, where the average value is taken over the steady state of the system. Under the weak-driving condition, the mean photon number is much less than $1$. Based on this, the condition $g^{(2)}\left(0\right)<1$ defining the sub-Poissonian photon-number statistics characterizes also single-PB (1PB), i.e., blockade of the subsequent photons by absorbing the first one~\cite{Scully1997Quantum,Testing2010Miranowicz}. Similarly, two-PB (2PB) fulfills the conditions $g^{(3)}\left(0\right)<1$ and $g^{(2)}\left(0\right)>1$ ~\cite{hamsen2017TwoPhoton}. For photon-induced tunneling (PIT), the absorption of the first photon favors also that of the second and subsequent photons, and it fulfills the conditions $g^{(\mu\geq2)}\left(0\right)>1$ corresponding to super-Poissonian photon-number statistics~\cite{faraon2008Coherent}. The correlation function can be calculated numerically by solving the quantum master equation [Eq.~(\ref{eq:MQ})].

In addition, we can use the effective Hamiltonian method to describe the evolution of the system. Here the evolution of the system is governed by the non-Hermitian Hamiltonian which is formed by adding phenomenologically the imaginary dissipation terms into Hamiltonian [Eq.~(\ref{eq:H})] as follows~\cite{Carmichael1993An,plenio1998quantum}
\begin{equation}\label{eq:He}
\hat{H}_{\mathrm{eff}}=\hat{H}-i\frac{\gamma}{2}\left(\hat{a}_{\mathrm{cw}}^{\dagger}\hat{a}_{\mathrm{cw}}+\hat{a}_{\mathrm{ccw}}^{\dagger}\hat{a}_{\mathrm{ccw}}\right).
\end{equation}
Under the weak-driving condition ($\xi/\gamma\ll1$), the Hilbert space can be restricted within a subspace with few photons. In the subspace with $N=m+n=3$ excitations, a general state of the system can be expressed as
\begin{equation}
|\psi(t)\rangle =\sum\limits_{N=0}^{3}\sum\limits_{m=0}^{N}C_{m,N-m}|m,N-m\rangle,
\end{equation}
where $C_{mn}\left(t\right)$ are the probability amplitudes corresponding to the bare states $\left|m,n\right\rangle$. We substitute the above general state and the Hamiltonian given in Eq.~(\ref{eq:He}) into the Schr\"{o}dinger equation $i|\dot{\psi}\left(t\right)\rangle =\hat{H}_{\mathrm{eff}}\left|\psi\left(t\right)\right\rangle$.  In the weak-driving regime, we can approximate the probability amplitudes of the excitations as $C_{m,N-m}\sim(\xi/\gamma)^{N}$. By using a perturbation method and discarding higher-order terms in each equation for lower-order variables~\cite{Carmichael1991Quantum}, we obtain the following equations of motion for the probability amplitudes:
\begin{align}
i\dot{C}_{10}\left(t\right) & =\Delta_{1}C_{10}\left(t\right)+JC_{01}\left(t\right)+\xi C_{00}\left(t\right),\nonumber \\
i\dot{C}_{01}\left(t\right) & =\Delta_{2}C_{01}\left(t\right)+JC_{10}\left(t\right),\nonumber \\
i\dot{C}_{20}\left(t\right) & =\Delta_{3}C_{20}\left(t\right)+\sqrt{2}JC_{11}\left(t\right)+\sqrt{2}\xi C_{10}\left(t\right),\nonumber \\
i\dot{C}_{11}\left(t\right) & =\Delta_{4}C_{11}\left(t\right)+\sqrt{2}JC_{20}\left(t\right)+\sqrt{2}JC_{02}\left(t\right)+\xi C_{01}\left(t\right),\nonumber \\
i\dot{C}_{02}\left(t\right) & =\Delta_{5}C_{02}\left(t\right)+\sqrt{2}JC_{11}\left(t\right),\nonumber \\
i\dot{C}_{30}\left(t\right) & =\Delta_{6}C_{30}\left(t\right)+\sqrt{3}JC_{21}\left(t\right)+\sqrt{3}\xi C_{20}\left(t\right),\nonumber \\
i\dot{C}_{21}\left(t\right) & =\Delta_{7}C_{21}\left(t\right)+\sqrt{3}JC_{30}\left(t\right)+2JC_{12}\left(t\right)+\sqrt{2}\xi C_{11}\left(t\right),\nonumber \\
i\dot{C}_{12}\left(t\right) & =\Delta_{8}C_{12}\left(t\right)+2JC_{21}\left(t\right)+\sqrt{3}JC_{03}\left(t\right)+\xi C_{02}\left(t\right),\nonumber \\
i\dot{C}_{03}\left(t\right) & =\Delta_{9}C_{03}\left(t\right)+\sqrt{3}JC_{12}\left(t\right),
\end{align}
where
\begin{align}
\Delta_{1}&=\Delta_{0}+\Delta_{\mathrm{sag}}-i\gamma/2, \quad \Delta_{2}=\Delta_{0}-\Delta_{\mathrm{sag}}-i\gamma/2, \quad \Delta_{3}=2\left(\Delta_{1}+\chi\right), \nonumber \\ \Delta_{4}&=\Delta_{1}+\Delta_{2}+2\chi, \qquad\;\: \Delta_{5}=2\left(\Delta_{2}+\chi\right), \qquad \quad\;\;\: \Delta_{6}=3\left(\Delta_{1}+2\chi\right), \nonumber \\ \Delta_{7}&=2\Delta_{1}+\Delta_{2}+6\chi, \qquad \Delta_{8}=\Delta_{1}+2\Delta_{2}+6\chi, \qquad \Delta_{9}=3\left[\Delta_{2}+2\chi\right].
\end{align}

\begin{figure*}[t]
    \centering
    \includegraphics[width=0.96\columnwidth]{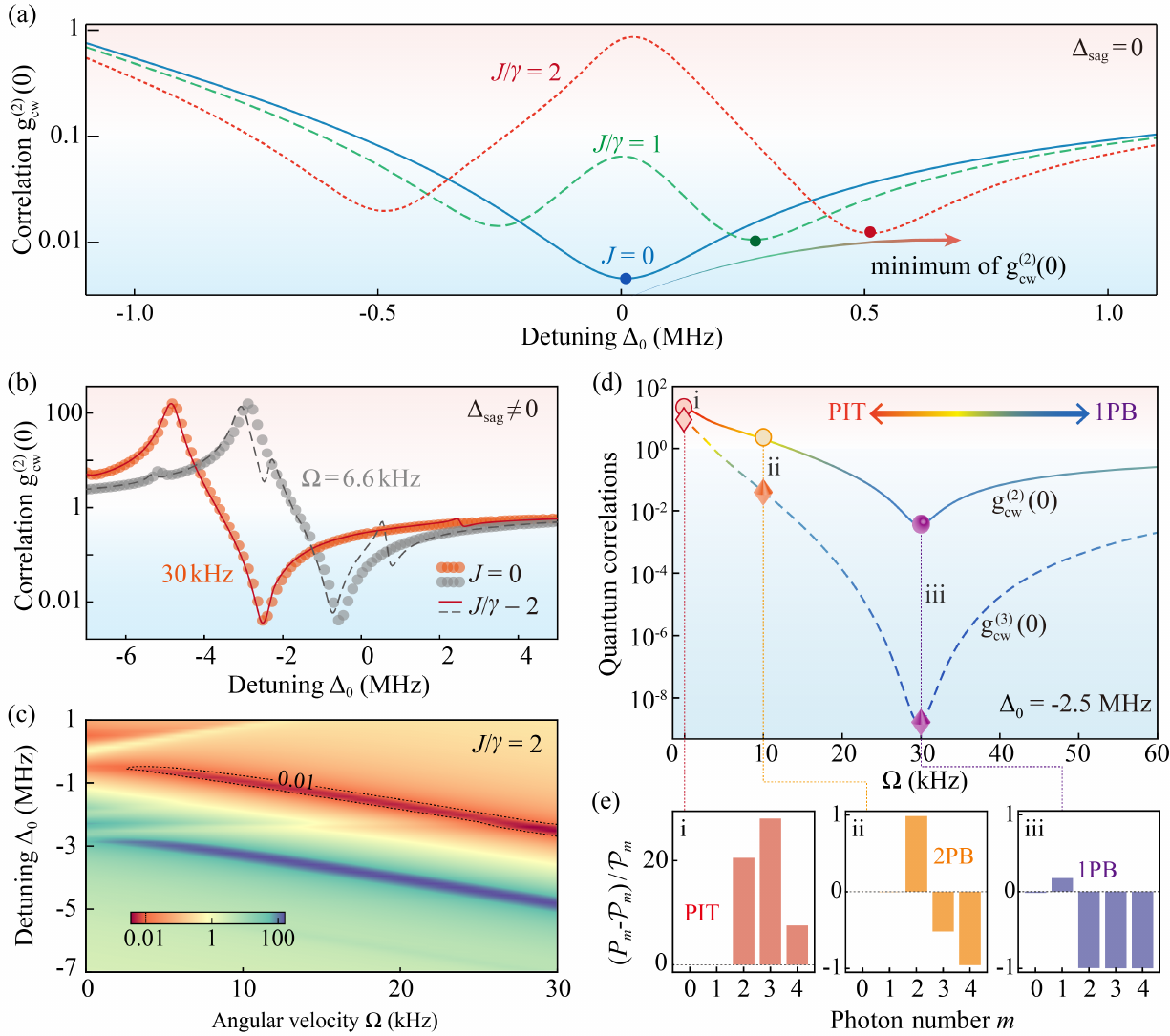}
    \caption{{Spinning-induced revival and a switch of photon blockade.} {(a)} The second-order correlation function $g_{\mathrm{\mathrm{cw}}}^{(2)}(0)$ versus the optical detuning $\Delta_{0}$ for different values of the coupling strength $J$. {(b)} $g_{\mathrm{cw}}^{(2)}(0)$ versus $\Delta_{0}$ for the cases with ($J/\gamma=2$, solid curves) and without backscattering ($J=0$, dotted curves) for different angular velocities $\Omega$. {(b)} The evolution of $g_{\mathrm{cw}}^{(2)}(0)$ as a function of $\Delta_{0}$ and $\Omega$. {(d)} The correlation functions $g_{\mathrm{cw}}^{(2)}(0)$ (solid curve) and $g_{\mathrm{cw}}^{(3)}(0)$ (dashed curve) versus $\Omega$. {(e)} This quantum switch can also be confirmed by comparing the photon-number distribution $P_{m}$ with the standard Poissonian distribution $\mathcal{P}_m$. The parameters are the same as those in Fig.~\ref{fig:FP1}.}
\label{fig:FP3}
\end{figure*}

By considering infinite-time limit condition ($t\rightarrow\infty$), the steady-state solutions of the probability amplitudes can be obtained
\begin{align}
C_{10} & =-\frac{\xi\Delta_{2}}{\eta_{1}},\quad C_{01}=\frac{J\xi}{\eta_{1}},\quad
C_{20}=\frac{-\sqrt{2}\xi\left(\eta_{3}C_{10}-J\Delta_{5}C_{01}\right)}{\varsigma_{1}},\nonumber \\
C_{11} & =\frac{\Delta_{5}\xi\left(2JC_{10}-\Delta_{3}C_{01}\right)}{\varsigma_{1}},\quad
C_{02} =\frac{\sqrt{2}J\xi\left(\Delta_{3}C_{01}-2JC_{10}\right)}{\varsigma_{1}},\nonumber \\
C_{30} & =\sqrt{3}\xi\frac{-\left(\eta_{5}\Delta_{7}-4J^{2}\Delta_{9}\right)C_{20}+\sqrt{2}J\eta_{5}C_{11}-2J^{2}\Delta_{9}C_{02}}{\varsigma_{2}},\nonumber \\
C_{21} & =\xi\frac{3J\eta_{5}C_{20}-\sqrt{2}\eta_{5}\Delta_{6}C_{11}+2J\Delta_{6}\Delta_{9}C_{02}}{\varsigma_{2}},\nonumber \\
C_{12} & =\xi\Delta_{9}\frac{-6J^{2}C_{20}+2\sqrt{2}J\Delta_{6}C_{11}-\eta_{4}C_{02}}{\varsigma_{2}},\nonumber \\
C_{03} & =\sqrt{3}\xi J\frac{6J^{2}C_{20}-2\sqrt{2}J\Delta_{6}C_{11}+\eta_{4}C_{02}}{\varsigma_{2}},
\end{align}
where $\eta_{1}=\Delta_{1}\Delta_{2}-J^{2}$, $\eta_{2}=\Delta_{3}\Delta_{4}-2J^{2}$, $\eta_{3}=\Delta_{4}\Delta_{5}-2J^{2}$, $\eta_{4}=\Delta_{6}\Delta_{7}-3J^{2}$, $\eta_{5}=\Delta_{8}\Delta_{9}-3J^{2}$,  $\varsigma_{1}=\Delta_{5}\eta_{2}-2J^{2}\Delta_{3}$, $\varsigma_{2}=\eta_{4}\eta_{5}-4J^{2}\Delta_{6}\Delta_{9}$. The probabilities of finding $m$ particles in the CW mode and $n$ particles in the CCW mode are given by
\begin{align}
P_{mn}=\left|C_{mn}\right|^{2}.
\end{align}
\begin{figure}[t]
    \centering
    \includegraphics[width=1.0\columnwidth]{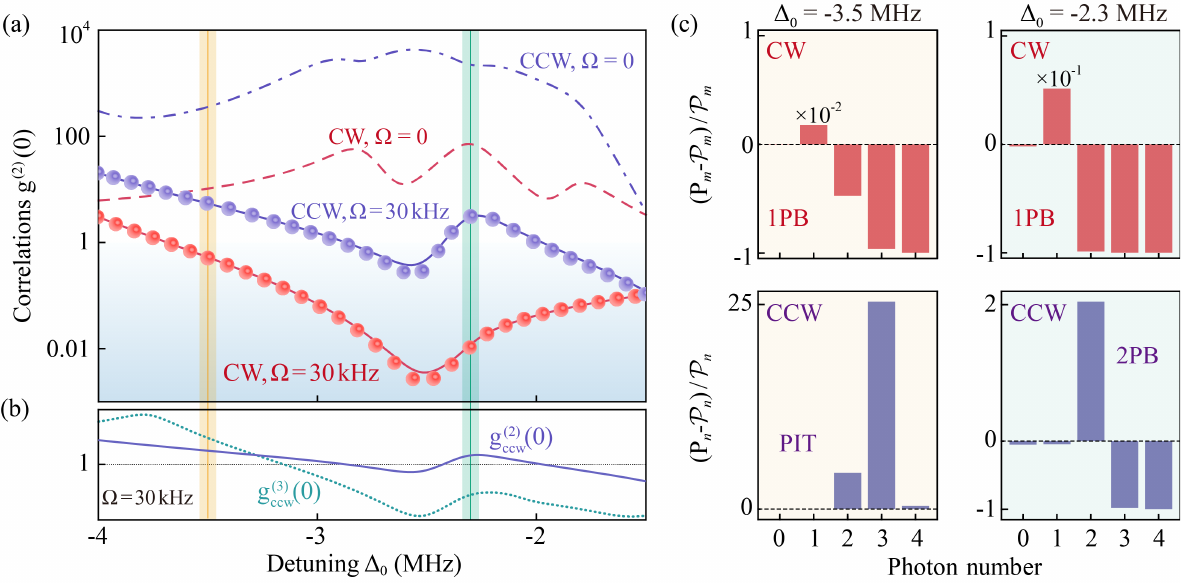}
    \caption{{Chiral photon blockade.} {(a)} The correlation functions $g_{\mathrm{cw}}^{(2)}\left(0\right)$ and $g_{\mathrm{ccw}}^{(2)}\left(0\right)$ versus the optical detuning $\Delta_{0}$ for the static (dashed curve) and spinning resonator (solid curves), where markers (circles) are analytical solutions of the spinning resonator case. Different quantum features appear in the CW and CCW modes, respectively. {(b)} The $g_{\mathrm{ccw}}^{(2)}\left(0\right)$ and $g_{\mathrm{ccw}}^{(3)}\left(0\right)$ versus $\Delta_{0}$ for $\Omega=30\:\mathrm{kHz}$. {(c)} The chiral photon blockade can also be recognized from the deviations of the photon-number distribution $P_{k}=\mathrm{Tr}[\left|k\right\rangle\left\langle k\right|\rho_{\mathrm{ss}}]~(k=m,n)$ to the standard Poissonian distribution $\mathcal{P}_{k}=\langle k\rangle^{k}\mathrm{exp}(-\langle k\rangle)/k!$ with the same mean photon number. The parameters are the same as those in Fig.~\ref{fig:FP1}.}
\label{fig:FP2}
\end{figure}
Based on this, we can finally derive the results for the second-order correlation functions:
\begin{align}
&g_{\mathrm{cw}}^{(2)}\left(0\right)\simeq\frac{2P_{20}}{P_{10}^{2}}=4\left|\eta_{1}\frac{\Delta_{2}\Delta_{4}\Delta_{5}+2J^{2}\chi}{\varsigma_{1}\Delta_{2}^{2}}\right|^{2},\label{eq:g2cw}\\
&g_{\mathrm{ccw}}^{(2)}\left(0\right)\simeq\frac{2P_{02}}{P_{01}^{2}}=16\left|\frac{\eta_{1}J^{2}\left(\Delta_{4}-\chi\right)}{\varsigma_{1}J^{2}}\right|^{2}.\label{eq:g2ccw}
\end{align}
In the absence of optical backscattering ($J=0$), the second-order correlation function~[as done in Eq.~(\ref{eq:g2cw})] of CW mode can be approximately written as
\begin{align}
g_{\mathrm{cw}}^{(2)}\left(0\right)\simeq\frac{\left(\Delta_{0}+\Delta_{\mathrm{sag}}\right)^{2}+\gamma^{2}/4}{\left(\Delta_{0}+\Delta_{\mathrm{sag}}+\chi\right)^{2}+\gamma^{2}/4},
\end{align}
which is consistent with a single-mode spinning resonator~\cite{huang2018Nonreciprocal}, as shown in the Fig.~\ref{fig:FP3}(b). While for the CCW mode, the $g_{\mathrm{ccw}}^{(2)}\left(0\right)$ in Eq.~(\ref{eq:g2ccw}) is meaningless due to $P_{02}=0$ and $P_{01}=0$.

In Fig.\,\ref{fig:FP2}, we plot $g_{\mathrm{cw}}^{(2)}\left(0\right)$ and $g_{\mathrm{ccw}}^{(2)}\left(0\right)$ as a function of the optical detuning $\Delta_{0}$ for the static case and the spinning case, respectively. It is seen that there is an excellent agreement between the analytical results (based on the semiclassical quantum-jump-free approach using the non-Hermitian Hamiltonian) and the numerical results (based on the fully quantum approach using the master equation). This result indicates that the effect of quantum jumps for the observation of PB can effectively be ignored for the studied ranges of the system parameters. We note that for some specific parameters, the effect of such quantum jump plays a key role in the evolution of the system dynamics and cannot be ignored in this situation~\cite{Minganti2019quantum}.

Figure \ref{fig:FP2}(a) shows that, for the static case, the second-order correlation functions always have the same photon-number statistics regardless of the direction of light propagation, that is, exhibiting $g_{\mathrm{cw,ccw}}^{(2)}\left(0\right)>1$ for $\Delta_{0}<-1.5\:\mathrm{MHz}$, which corresponds to the super-Poissonian photon-number statistics. In contrast, for the spinning case, the quantum features of the CW and CCW modes become different due to the spinning-induced chirality. We find that 1PB emerges around $\Delta_{0}=-3.5\:\mathrm{MHz}$ for the CW mode, while we have PIT for the CCW mode, i.e., $g_{\mathrm{cw}}^{(2)}\left(0\right)<1$ and $g_{\mathrm{ccw}}^{(2,3)}\left(0\right)>1$ [Figs.~\ref{fig:FP2}(a) and \ref{fig:FP2}(b)]. In addition, at $\Delta_{0}=-2.3\:\mathrm{MHz}$, 1PB occurs in the CW mode, due to $g_{\mathrm{cw}}^{(2)}\left(0\right)\sim0.01$, while 2PB occurs in the CCW mode [$g_{\mathrm{ccw}}^{(2)}\left(0\right)\sim3.04$, $g_{\mathrm{ccw}}^{(3)}\left(0\right)\sim0.02$]. These results can also be confirmed by comparing the photon-number distribution $P_{m}$ ($P_{n}$) with the Poissonian distribution $\mathcal{P}_{m}$ ($\mathcal{P}_{n}$) of the CW (CCW) mode [Fig.~\ref{fig:FP2}(c)]. We find that, for the CW mode, the single-photon probability $P_{1}$ is enhanced while $P_{m>1}$ are suppressed at $\Delta=-3.5\:\mathrm{MHz}$, which is in sharp contrast to the case of the CCW mode. These results reveal a direction-dependent quantum effect, i.e., chiral PB. This occurs due to the interplay of both the rotation-induced Sagnac effect and the nonlinearity-induced anharmonicity. For the CW mode, when it is driven by light that satisfies the single-photon resonance, photon blockade occurs due to the nonlinearity-induced anharmonicity of energy levels [$g_{\mathrm{cw}}^{(2)}\left(0\right)<1$]. As for the CCW mode, the Sagnac effect causes the resonance frequency of this mode to shift downward with increasing angular velocity. If the driving frequency remains unchanged, this shift of the resonance frequency can cause the original driving frequency, which does not satisfy the resonance condition, to match the two-photon resonance frequency of the CCW mode. Therefore, we can observe the result of $g_{\mathrm{ccw}}^{(2)}\left(0\right)>1$.

More interestingly, this spinning-induced chirality provides a feasible way to protect the devices against backscattering losses. Figure \ref{fig:FP3}(a) shows that, for a static cavity ($\Omega=0$), the minimum value of $g_{\mathrm{cw}}^{(2)}\left(0\right)$ increases with the increase of the backscattering-induced mode coupling rate $J$, resulting in the suppression of the emergence of PB. For the spinning resonator ($\Omega\neq0$), we find that the spinning-induced chirality has a revival effect on PB under the same non-ideal conditions ($J\neq0$), i.e., the minimum value of $g_{\mathrm{cw}}^{(2)}\left(0\right)$ significantly decreases [Fig.~\ref{fig:FP3}(b)]. For example, by choosing the angular velocity $\Omega=30\:\mathrm{kHz}$, we have $\mathrm{min}~[g_{\mathrm{cw}}^{(2)}\left(0\right)]\sim0.004$, which is almost the same as the case of an ideal cavity under the same conditions. Meanwhile, PB can gradually revive to the level of the ideal cavity with the increase of the angular velocity [Fig.~\ref{fig:FP3}(c)]. Therefore, we conclude that spinning-induced chirality has the ability to compensate or even counteract the negative effects of backscattering~\cite{Jiao2020Nonreciprocal}.

Figure \ref{fig:FP3}(d) shows different quantum effects, i.e., 1PB, 2PB, and PIT, which can be observed by tuning the angular velocity $\Omega$ for the case of non-resonance. We find that PIT emerges with $g_{\mathrm{cw}}^{(2)}\left(0\right)\sim21.55$ and $g_{\mathrm{cw}}^{(3)}\left(0\right)\sim8.58$ at $\Omega=0$ [Fig.~\ref{fig:FP3}(d)]. Increasing $\Omega$ to $10\:\mathrm{kHz}$, the correlation functions fulfill the conditions of the 2PB [$g_{\mathrm{cw}}^{(2)}\left(0\right)\sim1.98$ and $g_{\mathrm{cw}}^{(3)}\left(0\right)\sim0.04$]. By further increasing the angular velocity ($\Omega=30\:\mathrm{kHz}$), 1PB appears. This result can also be clearly seen in Fig.~\ref{fig:FP3}(e). With such a device, the switching between 1PB, 2PB, and PIT can be achieved by tuning the angular velocity.

\section{Conclusions}

In summary, we studied a novel quantum effect, the chiral photon blockade effect, extending previous work on the spontaneous breaking of classical optical chiral symmetry to the quantum domain. Specifically, we find that for a fixed driving field direction, the light propagating along the opposite direction in the microcavity exhibits completely different quantum effects, such as photon bunching and anti-bunching effects, due to the spinning-induced breaking of time-reversal symmetry. This novel chiral quantum device based on a single cavity has potential applications in chiral quantum networks~\cite{Pichler2015Quantum,Mahmoodian2016Quantum,Ramos2016Non,Mok2020Microresonators}. The spinning-cavity scheme is highly experimentally feasible and has been used to realize nonreciprocal optical transmission~\cite{Maayani2018Flying}. Significantly, the nonreciprocal quantum effect~\cite{huang2018Nonreciprocal} based on the spinning-cavity scheme has been experimentally demonstrated using different systems, which indicates that this novel effect in our work is also expected to be realized in more experimental systems, including cavity QED systems~\cite{Non2023Yang} and solid-state devices~\cite{Nonreciprocity2022Graf}. More interestingly, we have shown how to revive the photon blockade effect suppressed by backscattering, and switch different types of quantum correlations with the help of spinning which results can be further applied in chiral multi-photon bundles~\cite{Munoz2014Emitters,Bin2020NPhonon,Bin2021Parity} and backaction-immune quantum sensing~\cite{Stannigel2012Optomechanical,Mal2017Proposal,Fleury2015An,Liu2023Phase,Yang2015Invisible,Zhao2020Weak}.

\begin{backmatter}
\bmsection{Funding} National Natural Science Foundation of China (11935006, 12064010, 12147156, 12175060, 12247105); Science and Technology Innovation Program of Hunan Province (2020RC4047, 2022RC1203); Hunan Provincial Major Sci-Tech Program (2023ZJ1010); National Key and Development Program of China (2024YFE0102400); Narodowe Centrum Nauki (NCN) under the Maestro (DEC-2019/34/A/ST2/00081); XJ-Lab Key Project (23XJ02001).

\bmsection{Acknowledgments} The authors thank Ran Huang and Baijun Li for helpful discussions.

\bmsection{Disclosures} The authors declare no conflicts of interest.

\bmsection{Data availability} Data underlying the results presented in this paper are not publicly available at this time but may be obtained from the authors upon reasonable request.

\end{backmatter}

\end{document}